\documentstyle[preprint,epsfig,aps]{revtex}

\tightenlines

\begin{document}

\title{ Proton-neutron pairing in the deformed BCS approach}

\author{ 
F. \v Simkovic$^{1,2)}$, Ch.C. Moustakidis$^{1)}$,
L. Pacearescu$^{1)}$,
and Amand Faessler$^{1)}$
}

\address{
$^1$  Institute of Theoretical Physics, University of Tuebingen,
D--72076 Tuebingen, Germany\\
$^2$Department of Nuclear Physics,  
Comenius University, 
SK--842 15 Bratislava, Slovakia}
\date{\today}

\maketitle

\draft

\begin{abstract}
{We examine isovector and isoscalar proton-neutron pairing 
correlations for the ground state of even-even $Ge$
isotopes with mass number A=64--76 within the deformed BCS
approach. For N=Z $^{64}_{32}Ge$ the BCS solution 
with only T=0 proton-neutron pairs is found. For other nuclear systems ($N>Z$) 
a coexistence of a T=0 and T=1 pairs in the BCS wave function is observed. 
A problem of fixing of strengths of isoscalar and isovector 
pairing interactions is addressed. A dependence of number of
like and unlike pairs in the BCS ground state on the difference
between number of neutrons and protons is discussed.
We found that for nuclei with N much bigger than Z the effect
of proton-neutron pairing is small but not negligible.
}
\\
{PACS number(s): 21.60.-n, 21.60.Fw, 21.30.Fe, 23.40.Hc}
\end{abstract}

\section{Introduction}

The proton--neutron (pn) pairing correlations remain  to be subject of
a great interest as it is expected that they play an important role in
nuclear structure and decay for proton-rich nuclei with $N\approx Z$.
In these nuclei  proton and neutrons occupy identical orbitals and
have maximal spatial overlap. New experimental facilities
involving radioactive nuclear beams offer opportunities to study
$N=Z$ nuclei up to $^{100}Sn$ \cite{Grawe95,Grawe98}. There is still
much to be learned about systems out of the region of stability.
New knowledges could be helpful in understanding of various phases
of stellar evolution including nucleosynthesis and the abundance
of elements. Decay properties and nuclear structure are closely related.
The influence of the pn-pairing on the position and stability
of the proton drip line due to the additional pn-pairing
binding energy are becoming important issue in nuclear structure
\cite{Dobaczewski94-a,Nazarewicz94}. The recent progress in sensitivity
achieved with the large $\gamma$-ray detector arrays allows to study
the consequences of the pn pair correlations for the rotational spectra
\cite{rotat}.

The pn--pair correlations have been a major challenge to the nuclear
structure models for a long time (for a review of the early work on pn-pairing
problem see Ref. \cite{Goodman79}). In contrast to the proton--proton (pp)
and the neutron--neutron (nn) pairing, the proton--neutron pairing may exist
in two different varieties, namely isoscalar (T=0) and isovector  (T=1) pairing.
A generalized pairing formalism, which includes T=0 and T=1 pn-correlations, was
derived by Chen and Goswami \cite{Chen67}. The interplay of isovector and
isoscalar pairing has been studied in various contexts especially for N=Z nuclei
\cite{Goswami64,Chen67,camiz,Goodman68,Goodman70,wolter,Goodman72,Sandhu75,Sandhu76,Muller82}.
In recent publications phenomena like possible phase transition between different pairing
modes, competition of isoscalar and isovector pn-pairing and the ground state properties
of both even-even and odd-odd N=Z nuclei
 were studied mostly within schematic models
\cite{Engels97,Satula97,Goodman98,Satula00,Palchikov01,Engels96,Civitarese97,Engels98,Dobes97,martinez,polon}.
It was also shown that both single and double beta decay transitions
are affected by the proton-neutron pairing \cite{Cheoun95,Pantis96}.

The aim of this paper is to study the pn-pairing effect within generalized BCS approach
with schematic forces by taking into account the deformation degrees of freedom.
The main point is to use the advantage of the formalism constructed by
Chen and Goswami \cite{Chen67}, which is flexible
enough to account for both the T=1 and T=0 pairing correlations between nucleons
in time-reversed orbitals, in order to study  
the interplay and competition of isovector and isoscalar pairing. For that purpose
a schematic nuclear Hamiltonian with separated pp, nn and pn (T=1 and T=0)
pairing interactions is written.
We focus our attention also on the problem
whether the pn--pairing correlations are restricted only to the vicinity of the $N=Z$
line for medium heavy nuclei. Questions related to the fixing of pairing strength
parameters will be discussed.

\section{Theory}

The ground state of even-even nuclei is determined by the
deformed pairing Hamiltonian, which  includes monopole ($K=0$)
proton, neutron and proton--neutron pairing
interactions:
\begin{eqnarray}
H&=&
\sum_{s}(\epsilon_{p s}^{0}-\lambda_p)
\sum_{\sigma} c_{p s\sigma}^{\dagger}   c_{ps \sigma}+
\sum_{s}(\epsilon_{n s}^{0}-\lambda_n)
\sum_{\sigma} c_{ns\sigma}^{\dagger}   c_{ns \sigma}
\nonumber\\
&&-G_{pp}^{T=1} \sum_{s,s'} {S^{T=1}_{spp}}^{\dagger} S^{T=1}_{s'pp}
  - G_{nn}^{T=1} \sum_{s,s'} {S^{T=1}_{snn}}^{\dagger} S^{T=1}_{s'nn}
\nonumber \\
&& - G_{pn}^{T=1} \sum_{s,s'} {S^{T=1}_{spn}}^{\dagger} S^{T=1}_{s'pn}
- G_{pn}^{T=0} \sum_{s,s'} {S^{T=0}_{spn}}^{\dagger} S^{T=0}_{s'pn},
\label{eq:1}
\end{eqnarray}
where the  $\epsilon_{p s}^{0}$ and $\epsilon_{n s}^{0}$ are
the unrenormalized proton and neutron single particle energies,
respectively.
$\lambda_p$ ($\lambda_n$) is the proton (neutron) Fermi energy and
${S^{T}_{s\tau \tau'}}^{\dagger}$  creates isovector (T=1) or
isoscalar (T=0) pairs in time reversed orbits \cite{polon}
\begin{eqnarray}
{S^{T=1}_{spp}}^{\dagger} &=& \sum_\sigma
c_{ps\sigma}^{\dagger} c_{ps \tilde{\sigma}}^{\dagger},
~~~
{S^{T=1}_{snn}}^{\dagger} = \sum_\sigma
c_{ns\sigma}^{\dagger} c_{ns \tilde{\sigma}}^{\dagger},
\nonumber\\
{S^{T=1}_{spn}}^{\dagger} &=& \sum_\sigma
\frac{1}{\sqrt{2}}
(c_{ps\sigma}^{\dagger} c_{ns \tilde{\sigma}}^{\dagger}+
c_{ns\sigma}^{\dagger} c_{ps \tilde{\sigma}}^{\dagger}),
\nonumber\\
{S^{T=0}_{spn}}^{\dagger} &=& \sum_\sigma
\frac{1}{\sqrt{2}}
(c_{ps\sigma}^{\dagger} c_{ns \tilde{\sigma}}^{\dagger}-
c_{ns\sigma}^{\dagger} c_{ps \tilde{\sigma}}^{\dagger}).
\label{eq:2}
\end{eqnarray}
Here, $c_{\tau s\sigma}^{\dagger}$ and $c_{\tau ps\sigma}$ stand for the
creation and annihilation operators of a particle ($\tau=p$ and $\tau=n$
denote proton and neutron, respectively) in the axially--symmetric
harmonic oscillator potential. These states are completely
determined by a principal set of quantum numbers
$s = (N,n_z,\Lambda,\Omega)$. $\sigma$ is the sign of the angular
momentum projection $\Omega$ ($\sigma=\pm 1$). We note that the intrinsic
states are twofold degenerate. The states with $\Omega$ and $-\Omega$
have the same energy as a consequence of the time reversal invariance.
$\sim$ indicates time reversed states.

The Hamiltonian in Eq.
(\ref{eq:1}) is invariant under hermitian and time reversal operations.
The four  coupling strengths $G^{T=1}_{pp}$, $G^{T=1}_{nn}$,
$G^{T=1}_{pn}$ and $G^{T=0}_{pn}$ are real and characterize
the associated isovector (pp, nn and pn) and
isoscalar (pn) monopole (K=0) pairing interactions. The isospin
symmetry of the  Hamiltonian in Eq. (\ref{eq:1}) is restored
for  $\epsilon_{p s}^{0} = \epsilon_{n s}^{0}$
and
$G^{T=1}_{pp} = G^{T=1}_{nn} = G^{T=1}_{pn} = G^{T=0}_{pn}$.
In particular case where $G^{T=1}_{pn} = G^{T=0}_{pn}$
we get
\begin{equation}
H = \sum_{s \sigma \tau} (\epsilon_{\tau s}^{0} - \lambda_\tau)
c_{\tau s\sigma}^{\dagger} c_{\tau s {\sigma}}
-
\sum_{\tau\tau'} G_{\tau\tau'} \sum_{s\sigma s'\sigma'}
c_{\tau s\sigma}^{\dagger} c_{\tau' s \tilde{\sigma}}^{\dagger}
c_{\tau' s' \tilde{\sigma}'} c_{\tau s' \sigma'}.
\label{eq:3}
\end{equation}
It is assumed that $G_{\tau\tau'} = G_{\tau'\tau}$. In this
limit one can not distinguish between T=0 and T=1 pairing.
We note that a similar Hamilton was discussed in Ref. \cite{Civitarese97},
where the representation of
the single particle states with good angular momentum quantum
number was considered.


If the proton-proton, neutron-neutron and proton-neutron pairing
correlations  are considered for axially-symmetric nuclei,
the particle ($c^{+}_{\tau s \sigma}$ and
$c^{}_{\tau s \sigma}$, $\tau = p,n$) and the quasiparticle
($a^{+}_{\rho s \sigma}$ and $a^{}_{\rho s \sigma}$,
$\rho = 1,2$) creation and annihilation operators for the
deformed shell model states are related each to other by the
generalized BCS  transformation \cite{Goodman68}:
\begin{equation}
\left( \matrix{ c^{+}_{p s \sigma } \cr
c^{+}_{n s \sigma } \cr {{c}}_{p s \tilde{\sigma} } \cr
{{c}}_{n s \tilde{\sigma}}
}\right) = \left( \matrix{
u_{s 1 p} & u_{s 2 p} & -v_{s 1 p} & -v_{s 2 p} \cr
u_{s 1 n} & u_{s 2 n} & -v_{s 1 n} & -v_{s 2 n} \cr
v_{s 1 p} & v_{s 2 p} & u_{s 1 p} & u_{s 2 p} \cr
v_{s 1 n} & v_{s 2 n} & u_{s 1 n} & u_{s 2 n} }\right)
\left( \matrix{ a^{+}_{1 s \sigma} \cr
a^{+}_{2 s {\sigma}} \cr {{a}}_{1 s \tilde{\sigma}} \cr
{{a}}_{2 s \tilde{\sigma}} }\right),
\label{eq:4}
\end{equation}
where the occupation amplitudes
$u_{s 1 p}$, $v_{s 1 p}$, $u_{s 2 n}$, $v_{s 2 n}$ are real
and  $u_{s 1 n}$, $v_{s 1 n}$, $u_{s 2 p}$, $v_{s 2 p}$ are
complex \cite{Goodman79}.
In the case  only the T=1 proton-neutron pairing is considered
all amplitudes are real \cite{Goodman79,Cheoun95}.
In the limit in which there is no proton-neutron pairing
$u^{}_{s 2 p}$ = $v^{}_{s 2 p}$ = $u^{}_{s 1 n}$ = $v^{}_{s 1 n}$ = 0.
Then the isospin generalized BCS  transformation in
Eq. (\ref{eq:4}) reduces  to two conventional BCS
two--dimensional transformations,
first for protons ($u^{}_{s 1 p}=u^{}_{s p}$,
$v^{}_{s 1 p}=v^{}_{s p}$) and second for neutrons
($u^{}_{s 2 n}=u^{}_{s n}$, $v^{}_{s 2 n}=v^{}_{s n}$).


The diagonalization of the Hamiltonian (\ref{eq:1}) is
equivalent to the matrix diagonalization \cite{Goodman68}
\begin{equation}
\left( \matrix{
\epsilon_{ps}-\lambda_{p} & 0  & \Delta_{pp} &
\Delta_{pn} \cr
 0 & \epsilon_{ns}-\lambda_{n} & \Delta_{pn}^{*} &
\Delta_{nn} \cr
\Delta_{pp} & \Delta_{pn}&
-(\epsilon_{ps}-\lambda_{p})  & 0 \cr
\Delta_{pn}^{*} & \Delta_{nn} &
0  & -(\epsilon_{ns}-\lambda_{n}) \cr  }\right)
\left(\matrix{u_{s \rho p}^{}\cr u_{s \rho n}^{} \cr
v_{s \rho p}\cr v_{s \rho n}}\right)
= E_{s \rho}
\left(\matrix{u_{s \rho p}^{}\cr u_{s \rho n}^{} \cr
v_{s \rho p}\cr v_{s \rho n}}\right)
\label{eq:5}
\end{equation}
that yields the quasiparticle energies $E_{s \rho}$
and the occupation amplitudes.
Here, $\epsilon_{\tau s}$ ($\tau = p,n$) are the renormalized
single particle energies which include terms describing
the coupling of the nuclear average field with the characteristics
of the pairing interactions \cite{solov}. The proton
($\Delta_{pp}$), neutron ($\Delta_{nn}$) and proton--neutron
($\Delta_{pn}$) pairing gaps are given as
\begin{eqnarray}
\Delta_{\tau\tau } &= & G_{\tau \tau}^{T=1}
\sum_{s, \rho} v_{s \rho \tau} u_{s \rho \tau}^{*}
= G_{\tau \tau}^{T=1}
\sum_{s, \rho} v^*_{s \rho \tau} u_{s \rho \tau} ~~(\tau~=~p,n),
\nonumber\\
\Delta_{p n} &= & \Delta^{T=1}_{p n} + i  \Delta^{T=0}_{p n}
\label{eq:6}
\end{eqnarray}
with
\begin{eqnarray}
\Delta^{T=1}_{p n} &=& G^{T=1}_{pn}~Re\{~
\sum_{s, \rho} v_{s \rho p} u^*_{s \rho n}~\},\nonumber \\
 \Delta^{T=0}_{p n} &=& G^{T=0}_{pn}~Im\{~
\sum_{s, \rho} v_{s \rho p} u^*_{s \rho n}~\}.
\label{eq:7}
\end{eqnarray}
The real and imaginary parts of the proton--neutron pairing gap
$\Delta_{pn}$ are associated with T=1 and T=0 pairing modes, respectively.
This phenomenon was first pointed out by Goswami \cite{Chen67,Goodman68}, 
what made possible almost all subsequent treatments of pn pairing.
We note that for $G^{T=0}_{pn}$ equal to zero the occupation amplitudes
of the isospin generalized BCS transformations are real.
The Langrange multipliers $\lambda_p$ and $\lambda_n$ entering (\ref{eq:5})
are adjusted so that the number-conservation relations
\begin{equation}
Z = 2 \sum_{s \rho} v_{s \rho p} v^*_{s \rho p}, ~~
N = 2 \sum_{s \rho} v_{s \rho n} v^*_{s \rho n}
\label{eq:8}
\end{equation}
are satisfied.

The ground state energy can be written as
\begin{equation}
H_{g.s.}=H_{0}+H_{pair.},
\label{eq:9}
\end{equation}
where  $H_{0}$ is the BCS expectation value of the single-particle
Hamiltonian
\begin{equation}
H_{0}= 2\sum_{\tau s}\epsilon_{\tau s}
\sum_{\rho} v_{s \rho \tau} v_{s \rho \tau}^{*}
\label{eq:10}
\end{equation}
and  $H_{pair.}$ represents the pairing energy
\begin{equation}
H_{pair}=
-\frac{\Delta^2_{pp}}{G^{T=1}_{pp}}
-\frac{\Delta^2_{nn}}{G^{T=1}_{nn}}
-\frac{(\Delta^{T=1}_{pn})^2}{G^{T=1}_{pn}}
-\frac{(\Delta^{T=0}_{pn})^2}{G^{T=0}_{pn}}.
\label{eq:11}
\end{equation}
We note that pp, nn and pn  (T=0 and T=1) pairing modes
contribute coherently to the ground state energy $H_{g.s.}$.

In Ref. \cite{Engels96} it has been suggested
that the effect of different pairing modes can be quantified by
measuring pair numbers in the nuclear wave function \cite{Dobes97}.
For that purpose we define the operators
\begin{eqnarray}
{\cal N}_{pp}&=&  \sum_{s,s'} {S^{T=1}_{spp}}^{\dagger} S^{T=1}_{s'pp}, ~~~
{\cal N}_{nn}=\sum_{s,s'}  {S^{T=1}_{snn}}^{\dagger} S^{T=1}_{s'nn},
\nonumber\\
{\cal N}_{pn}^{T=1}&=& \sum_{s,s'} {S^{T=1}_{spn}}^{\dagger} S^{T=1}_{s'pn}, ~~~
{\cal N}_{pn}^{T=0}= \sum_{s,s'}  {S^{T=0}_{spn}}^{\dagger} S^{T=0}_{s'pn},
\label{eq:12}
\end{eqnarray}
which are rough measures of the numbers pp, nn, pn (T=1) and pn (T=0) pairs,
respectively. The BCS ground state expectation values of these operators
are related with the corresponding pairing gaps. After subtracting the
mean field values we find
\begin{eqnarray}
<{\cal N}_{pp}> &\approx&  \frac{\Delta^2_{pp}}{(G^{T=1}_{pp})^2},~~~
<{\cal N}_{nn}> \approx \frac{\Delta^2_{nn}}{(G^{T=1}_{nn})^2},
\nonumber\\
<{\cal N}_{pn}^{T=1}> &\approx&
\frac{(\Delta^{T=1}_{pn})^2}{(G^{T=1}_{pn})^2}, ~~~
<{\cal N}_{pn}^{T=0}> \approx
\frac{(\Delta^{T=0}_{pn})^2}{(G^{T=0}_{pn})^2}.
\label{eq:13}
\end{eqnarray}
We note that the number of these pairs can not be observed directly.

\section{Empirical pairing gaps}

The magnitude of proton, neutron and proton--neutron pairing gaps
can be determined only indirectly from the experimental data.
Usually they are deduced from systematic study of
experimental odd-even mass differences:
\begin{eqnarray}
M(Z,N)_{odd-odd} &=& {\cal M}(Z,N) \nonumber\\
M(Z,N)_{odd-proton} &=& {\cal M}(Z,N) + \Delta_p^{emp.} \nonumber \\
M(Z,N)_{odd-neutron} &=& {\cal M}(Z,N) + \Delta_n^{emp.} \nonumber \\
M(Z,N)_{odd-odd} &=& {\cal M}(Z,N) + \Delta_p^{emp.}+
\Delta_n^{emp.} - \delta_{pn}^{emp.}.
\label{eq:14}
\end{eqnarray}
Here, $M(Z,N)$ are the experimental nuclear masses and
${\cal M}(Z,N)$
denotes a smooth mass surface formed by a set of
even--even nuclei, i.e. for these nuclei the measured
mass is identical to the smooth mass. The mass
of odd--proton (odd--neutron) nucleus is given by addition
of the proton pairing gap $\Delta_{p}^{emp.}$
(neutron pairing gap $\Delta_{n}^{emp.}$) to ${\cal M}(Z,N)$.
The mass of an odd-odd nucleus is the sum of the smooth mass
${\cal M}(Z,N)$
and the sum of the proton and neutron
pairing gaps minus the attractive residual proton--neutron
interaction energy $\delta_{pn}^{emp.}$.

Using  the Taylor series expansion of the
${\cal M}(Z,N)$, the quantities $\Delta_{p}^{emp.}$, $\Delta_{n}^{emp.}$
and  $\delta_{pn}^{emp.}$ for even mass nuclei
can be expressed as
\begin{eqnarray}
\Delta_{p}^{emp.} &=& -\frac{1}{8}[ M(Z+2,N)-4M(Z+1,N)+6 M(Z,N)
-4 M(Z-1,N)+M(Z-2,N)],\nonumber \\
\Delta_{n}^{emp.} &=& -\frac{1}{8}[ M(Z,N+2)-4M(Z,N+1)+6 M(Z,N)
-4 M(Z,N-1)+M(Z,N-2)],\nonumber \\
\delta_{pn}^{emp.} &=& \frac{1}{4} \{
2 [ M(Z,N+1) + M(Z,N-1)+ M(Z-1,N) + M(Z+1,N)] -4 M(Z,N)]
\nonumber\\
&& - [ M(Z+1,N+1) + M(Z-1,N+1) + M(Z+1,N-1) + M(Z-1,N-1)] \}
\nonumber \\
\label{eq:15}
\end{eqnarray}

The first systematic studies of nuclear masses have shown that
the average pairing gaps (${\overline{\Delta}}_{\tau\tau}$, $\tau = p,~n$)
decrease slowly with $A^{1/2}$
(traditional model) \cite{bohr}. Vogel $et~al.$ found evidence
for a dependence of the average pairing gaps upon the
relative neutron excess $(N-Z)/A$ \cite{vogl}. The parameterizations of
the average pairing gaps and the average proton-neutron residual
interaction within these two models are as follows:
\begin{eqnarray}
{\overline{\Delta}}_{\tau} &=& 12~MeV/A^{1/2}, ~~~~
{\overline{\delta}}_{pn} = 20~MeV/A~~(traditional~ model)\nonumber\\
{\overline{\Delta}}_{\tau} &=& (7.2-44\frac{(N-Z)^2}{A^2})~MeV/A^{1/3}, ~~~~
{\overline{\delta}}_{pn} = 31~MeV/A~~(Vogel~et.al).
\label{eq:16}
\end{eqnarray}
We note that recently Madland and Nix \cite{madl} presented a model for
calculation of these average quantities by fixing a
larger set of parameters.

\begin{figure}
\begin{center}
\begin{tabular}{cc}
{\epsfig{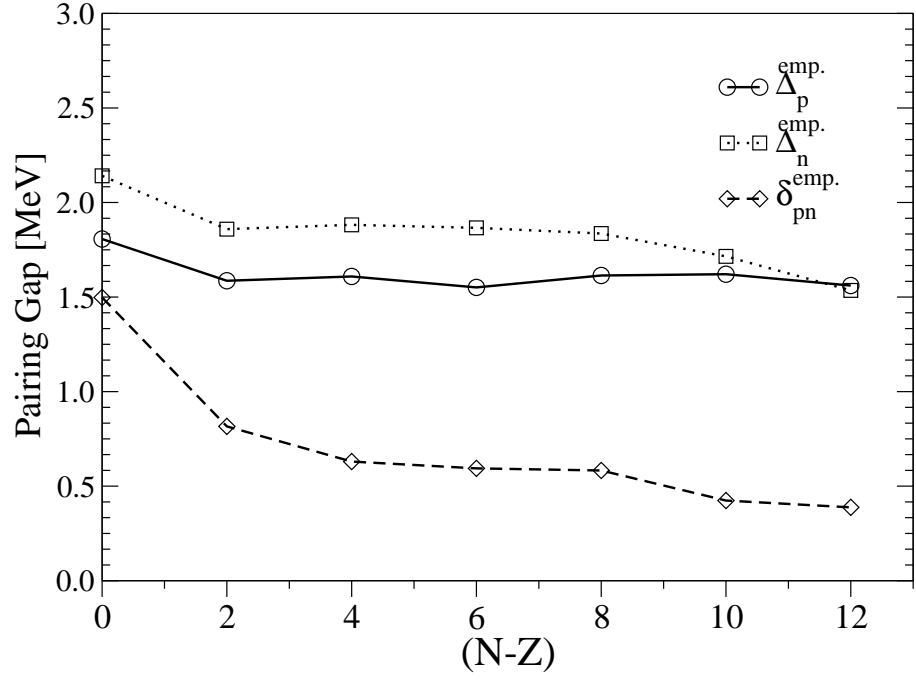} }
\end{tabular}
\end{center}
\caption{ The experimental proton ($\Delta^{emp.}_{p}$)
and neutron ($\Delta^{emp.}_{n}$) pairing gaps and
proton--neutron interaction energy ($\delta^{emp.}_{pn}$)
for even--even Ge isotopes
with A=64--76 [see Eq. (\protect\ref{eq:15})]}.
\label{fig.1}
\end{figure}

In Table \ref{table.1} we present
the calculated experimental pairing gaps and proton--neutron
excitation energies for Ge isotopes with A=64--76 and compare
them with the averaged ones. We see that a better agreement
between empirical and average values is achieved for
the model developed by  Vogel et al. \cite{vogl}.
The differences between empirical and average values
are small especially for isotopes close to the valley of $\beta$
stability. It is worthwhile
to notice that the values of proton--neutron interaction
energies are not negligible  in comparison with the values of
pairing gaps even  for isotopes with large neutron excess.
This fact is clearly illustrated in Fig. \ref{fig.1}.
Thus the proton--neutron pairing interaction is expecting to play
a significant role in construction of the quasiparticle
mean field even for these nuclei. It is supposed that
the origin of this phenomenon is associated with the effect
of nuclear deformation, which is changing the
distribution of proton and neutron single particle
levels inside the nucleus.

For performing a realistic calculation within the deformed
BCS approach it is necessary to fix the parameters of the
nuclear Hamiltonian in Eq. (\ref{eq:1}).
Following the procedure of Ref. \cite{Cheoun95} it is done in
two steps:\\
i) The proton (neutron) pairing interaction strength $G^{T=1}_{pp}$
($G^{T=1}_{nn}$) are adjusted by requesting that the lowest proton
(neutron) quasiparticle energy to be equal to the empirical
proton (neutron) pairing gap $\Delta^{emp.}_p$  ($\Delta^{emp.}_n$).\\
ii) With already fixed $G^{T=1}_{pp}$ and $G^{T=1}_{nn}$ we adjust
the proton--neutron pairing interaction strengths
$G^{T=1}_{pn}$ and $G^{T=0}_{pn}$  to the empirical
proton--neutron interaction energy $\delta^{emp.}_{pn}$
using the formula
\begin{equation}
\delta_{pn}^{theor.} = - [ (H^{(12)}_{g.s.} + E_1 + E_2)
- (H^{(pn)}_{g.s.} + E_p + E_n) ].
\label{eq:17}
\end{equation}
Here, $H_{g.s.}^{(12)}$ ($H_{g.s.}^{(pn)}$) is the total
deformed BCS ground state energy with (without)
proton-neutron pairing and $E_1 + E_2$ ($E_p + E_n$) is the
sum of the lowest two quasiparticles energies with
(without) proton--neutron pairing gap $\Delta_{pn}$.

We note that the calculation of ground state energies of odd-odd
nuclei within macroscopic pairing models is based
on the assumption that there are one unpaired proton and neutron
with energies close to the Fermi energies \cite{vogl,moll}. 
The resulting expectation 
value of an attractive short-range residual interaction between them, 
which can be  approximated by a delta force, is considered to be the 
origin of the proton-neutron  interaction energy.
Unfortunatelly, this simplified approach can not be exploited in
microscopic treatment of nuclear properties of open shell nuclei,
as the construction of the many-body wave function is required.

In our deformed BCS approach the ground state of the odd-odd
nucleus is described as the lowest two quasiparticle excitation of the
even-even nucleus. The considered procedure of fixing the pairing 
strengths were exploited already in Refs.  \cite{Cheoun95,Moller92}. 
However, some questions arise about the ambiguity of equating the 
pairing gap expressions that are used to determine the strength of 
pairing matrix elements for microscopic pairing calculations
with the macroscopic pairing-gap model that is used to describe
average mass differences. Thus, we shall study the importance of the
proton-neutron pairing effect for $N>Z$ nuclei also by assuming a 
different scenario, namely commonly used pairing strengths,
\begin{equation}
G^{T=1}_{pp} = G^{T=1}_{nn} = 16/A~MeV, ~~~~
G^{T=0}_{pn} = 20/A~MeV,
\label{eq:18}
\end{equation}
which decrease with increasing neutron excess.

\section{Results and Discussion}

The starting point of our calculations is the eigenstates of a
deformed axially-symmetric Woods-Saxon potential
with the parameterization  of Ref.
\cite{Tan79}, i.e., spherical symmetry is broken already from the beginning.
For description of the ground states  of
Ge isotopes we use the values of the quadrupole ($\beta_2$)
and the hexadecapole ($\beta_4$) nuclear deformation
parameters from Ref. \cite{ring},
which are in good agreement with the predictions of the
macroscopic-microscopic model of M\"oller, Nix, Myers and Swiatecki
\cite{moll}. In the BCS calculation the single particle states
are identified with the asymptotic quantum numbers
$(N,n_z,\Lambda,\Omega)$.
We note that intrinsic states are twofold degenerate. The
states with $\Omega$ and $-\Omega$ have the same energy as
consequence of the time reversal invariance.
A truncated model space
with $N\le 5$  is considered. As it was stated in the
Section II only the coupling of nucleon
states in time-reversed components of the same orbitals
are taken into account.

We performed calculations within generalized BCS formalism
associated with the nuclear Hamiltonian in Eq. (\ref{eq:1}).
The solutions obtained can be classified as follows:\\
i) The BCS solution without pn-pairing. In this case
$\Delta_{pp}$ and $\Delta_{nn}$ are real and $\Delta_{pn}$=0.\\
ii) The BCS solution with T=1 pn-pairing. It corresponds to
the case the  $\Delta_{pp}$, $\Delta_{nn}$ and $\Delta_{pn}$ are real
($\Delta^{T=0}_{pn}$=0), i.e., all the occupation amplitudes are real.\\
iii) The BCS solution with T=0 pn-pairing, which is characterized
by real $\Delta_{pp}$ and $\Delta_{nn}$ and purely imaginary $\Delta_{pn}$ 
($\Delta^{T=1}_{pn}$=0). In this case the occupation amplitudes 
associated with pn-pairing ($u_{s1n}$, $v_{s1n}$, $u_{s2p}$ and $v_{s2p}$) 
are imaginary.

No coexistence of $T=0$ and $T=1$ proton--neutron pairing modes were
found. There is a very simple competition between the two kinds of  pn-pairing.
For $G^{T=1}_{pn} > G^{T=0}_{pn}$ and $G^{T=1}_{pn} < G^{T=0}_{pn}$
scenarios ii) and iii) are realized, respectively. In a particular
case $G^{T=1}_{pn} = G^{T=0}_{pn}$ both T=0 and T=1 pairing modes are
indistinguishable as it was indicated in Section II. We note that
the absolute values of the occupation amplitudes associated with
the solutions ii) and iii) are equal one to another if the T=1 pn--pairing
strength used in generating ii) solution is equal to the T=0
pn--pairing strength considered in the calculation of iii) solution
(proton and neutron pairing strengths are the same). In the case of
 $N=Z$ ($^{64}Ge$)  for enough large pn-pairing strength $G^{T=0}_{pn}$
or $G^{T=1}_{pn}$ a BCS solution without like--particle pairing modes
was observed.

\begin{figure}
\begin{center}
\begin{tabular}{cc}
{\epsfig{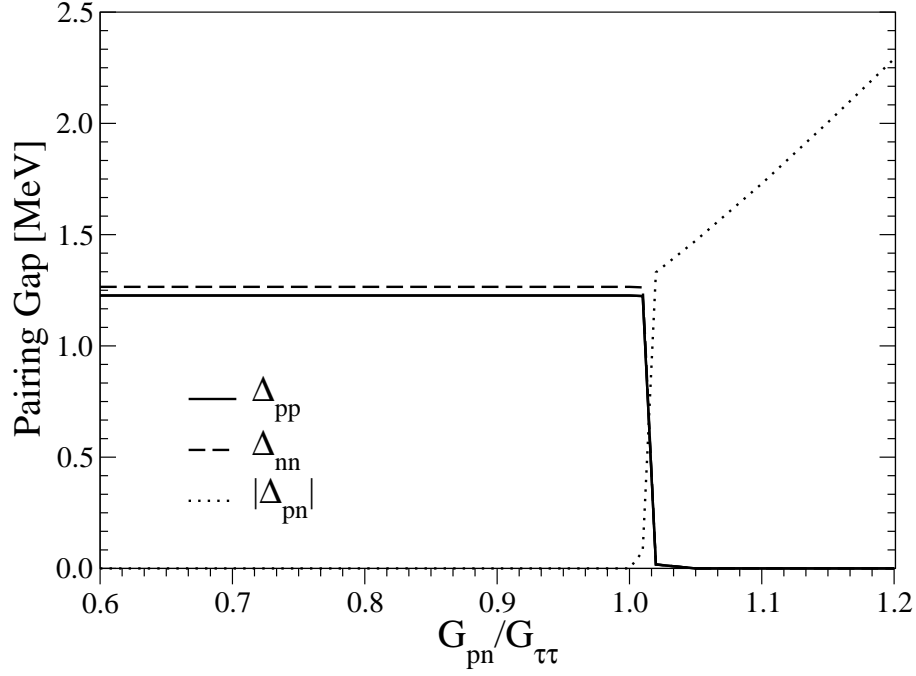} }
\end{tabular}
\end{center}
\caption{
The proton ($\Delta_{pp}$), neutron ($\Delta_{nn}$) and proton--neutron ($\Delta_{pn}$)
pairing gaps as a function of the ratio $G_{pn}/G_{\tau\tau}$ for the
$_{32}^{64}Ge$. $G_{\tau\tau}$ represents the proton and neutron pairing
strengths  ($G_{\tau\tau}=G^{T=1}_{pp}=G^{T=1}_{nn}$). $G_{pn}$ stands
for the larger of T=0 ($G^{T=0}_{pn}$) and T=1  ($G^{T=1}_{pn}$)
proton-neutron pairing strengths. $G_{\tau\tau}$ were taken to be $0.250~MeV$.
}
\label{fig.2}
\end{figure}

\begin{figure}
\begin{center}
\begin{tabular}{cc}
{\epsfig{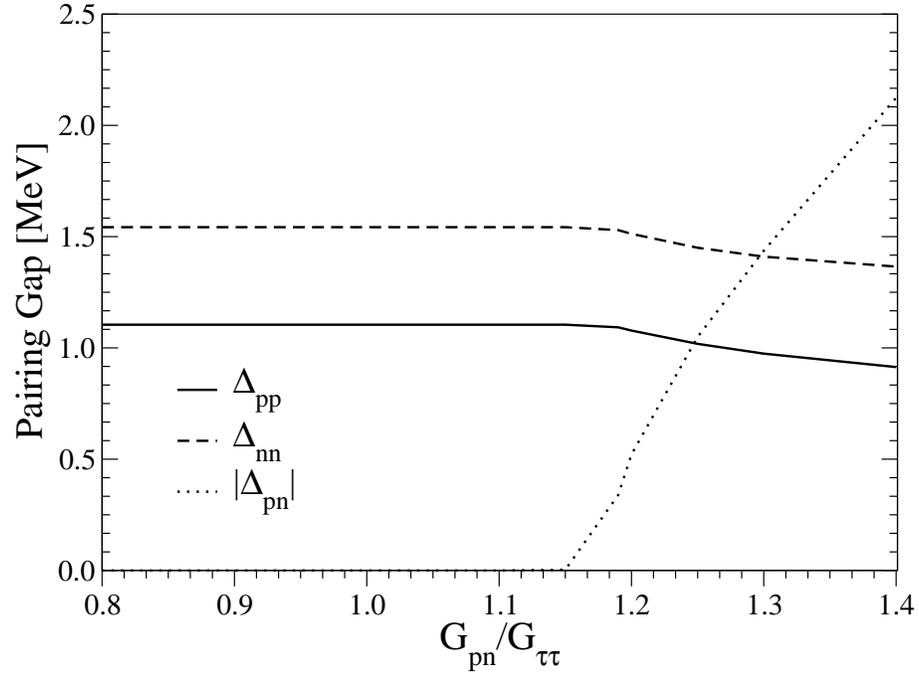} }
\end{tabular}
\end{center}
\caption{ 
The proton ($\Delta_p$), neutron ($\Delta_n$) and proton--neutron ($\Delta_{pn}$) 
pairing gaps as a function of the ratio $G_{pn}/G_{\tau\tau}$ for the
$_{32}^{70}Ge$. Conventions are the same as in Fig. \protect\ref{fig.2} and
$G_{\tau\tau}$ were equal to $0.229~MeV$.}
\label{fig.3}
\end{figure}

In Figs. \ref{fig.2} and \ref{fig.3} we show the BCS gap parameters as a function of
the ratio $G_{pn}/G_{\tau\tau}$ for $^{64}Ge$ and $^{70}Ge$, respectively.
The $G_{pn}$ stands for the larger of the T=1 ($G^{T=1}_{pn}$)
and T=0 ($G^{T=0}_{pn}$) proton--neutron pairing strengths and
$G_{\tau\tau}=G^{T=1}_{pp}=G^{T=1}_{nn}$. We stress that there is 
no coexistence of $T=0$ and $T=1$ proton--neutron pairing modes and that 
the absolute value of the pn-pairing gap $\Delta_{pn}$ is the same 
in the case of $T=1$ ($G_{pn}=G^{T=1}_{pn}>G^{T=0}_{pn}$)
and $T=0$ ($G_{pn}=G^{T=0}_{pn}>G^{T=1}_{pn}$) pairing solutions. 
In the case of $^{64}Ge$ ($^{70}Ge$), the  $G_{\tau\tau}$ was assumed  to be
$0.250$ MeV ($0.229$ MeV). Below some critical value of
$G_{pn}/G_{\tau\tau}$  there are only
proton and neutron pairing modes. For $^{64}Ge$
there is  only a narrow region above this critical
point in which like--particle and
proton--neutron pairs coexist. With additional  increase of the ratio
$G_{pn}/G_{\tau\tau}$ the system prefers to form only proton--neutron
pairs.  For nuclei with non-zero neutron excess ($N\not=Z$) like
$^{70}Ge$  there is a different situation.
In Fig. \ref{fig.3} we notice a  less sharp phase transition to the
proton-neutron pairing mode in comparison with that in Fig. \ref{fig.2}.
In addition, the proton-neutron pairing mode does exist only in
coexistence with the like particle pairing modes.

The binding energy gains between a system with no proton-neutron
interaction and the system where proton--neutron pairs do exist.
The ground state energy decreases monotonically with increasing
$G^{T=0,1}_{pn}$. Although the energy gain due to pairing correlations
is rather modest, it is expected that pn correlations influence
many properties of the atomic nuclei. In order to perform corresponding
studies the problem of fixing  the pairing strength parameters
has to be understood.

\begin{figure}
\begin{center}
\begin{tabular}{cc}
{\epsfig{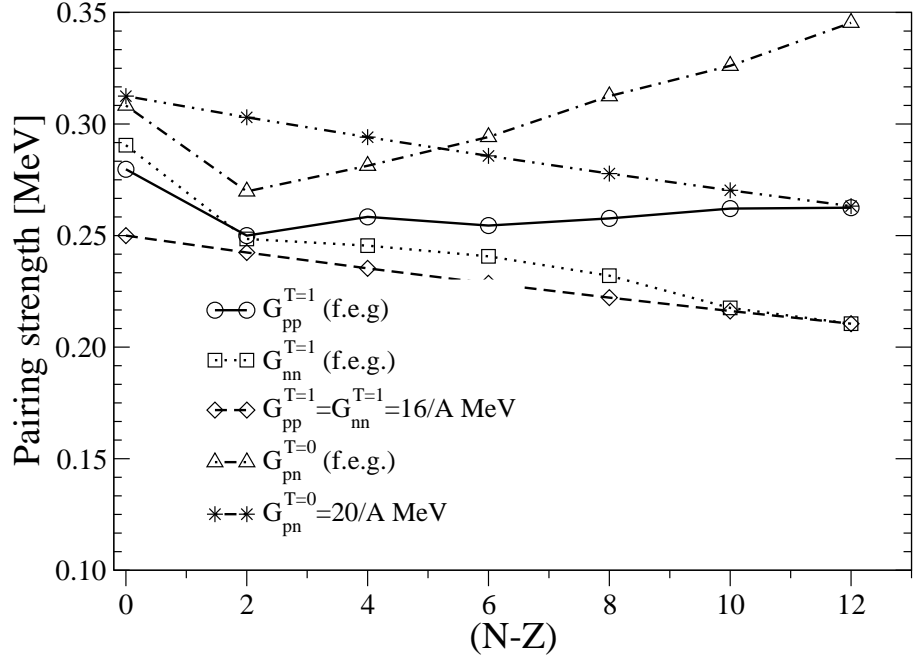} }
\end{tabular}
\end{center}
\caption{ The proton ($G^{T=1}_{pp}$), neutron ($G^{T=1}_{nn}$) and
proton--neutron ($G^{T=0}_{pn}$) pairing strengths as function
of the neutron excess $N-Z$. For the curves f.e.g. 
(fitted to the experimental gaps) the 
 strength is adjusted to
the experimental pairing gap ($\Delta^{emp.}_{p}$ or $\Delta^{emp.}_{n}$)
or  proton--neutron interaction energy
($\delta^{emp.}_{pn}$).
}
\label{fig.4}
\end{figure}

There is very little known about the T=0 and T=1 strengths of the pn-pairing.
The T=0, $^{3}S$ pairing force is expected to be stronger in comparison
with T=1, $^{1}S$
pairing forces. A strong evidence of it is that the deuteron and many
other double  even N=Z nuclei prefer this type of coupling
due to the strong tensor force contribution.
This fact favors the iii) solution in comparison with the ii) one.
In Fig. \ref{fig.4} the values of pairing strength adjusted to
experimental pairing gaps and proton-neutron interaction energy
(see previous Section for details) are presented. By comparing
$G^{T=1}_{pp}$ and $G^{T=1}_{nn}$ strengths we see that
the isospin invariance is significantly violated especially
for isotopes with large neutron excess (N-Z). The T=0 proton--neutron
force $G^{T=0}_{pn}$ is larger in comparison with T=1 pp and nn
($G^{T=1}_{pp}$ and $G^{T=1}_{nn}$) forces for all
considered Ge isotopes. The N=Z $^{64}Ge$ seems to be a special case.
For other Ge isotopes $G^{T=1}_{pp}$ is more or less stable with respect to the
N-Z difference and $G^{T=1}_{nn}$ slightly decreases
with increasing N-Z. The T=0 pn-force offers different scenario, namely,
$G^{T=0}_{pn}$ is slightly growing with increasing neutron excess $N-Z$,
what is surpricing. It can be due to the fact that only the
monopole pair Hamiltonian is considered within the deformed BCS
approach or connected with the way of adjusting it.
We note that the largest differences among
$G^{T=1}_{pp}$, $G^{T=1}_{nn}$ and   $G^{T=0}_{pn}$ forces are
visible for maximal value of $N-Z=12$. We note that for Ge isotopes
with $N-Z < 0$ and  $N-Z > 12$ the pairing strengths can not
be fixed following the procedure presented in the previous section
due to the lack of experimental information about
nuclear masses and/or proton and neutron separation energies.

We find it interesting to compare the behavior of the adjusted
pairing strengths with the commonly used prescriptions
for $G^{T=1}_{pp}$, $G^{T=1}_{nn}$ and   $G^{T=0}_{pn}$
[see Eq. (\ref{eq:18})]. From the Fig. \ref{fig.4} it is evident
that the agreement between them especially for $G^{T=1}_{pp}$ and
$G^{T=1}_{pn}$ forces is rather poor. The reason can be that
the considered strengths are expected to reproduce the general
behaviour througout the periodic table as function of A, but not
as neutron excess $N-Z$. Another possibilities are already
announced the simplicity of the considered nuclear model
or the limitations of the adjusting the parameters
of the microscopic pairing model to those of the macroscopic
pairing gap model.

\begin{figure}
\begin{center}
\begin{tabular}{cc}
{\epsfig{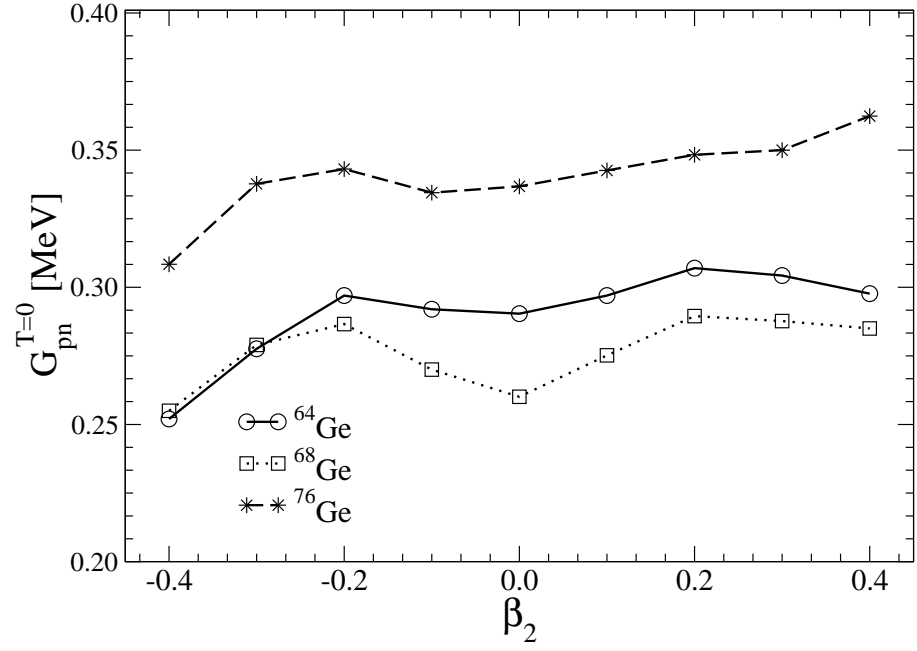} }
\end{tabular}
\end{center}
\caption{ The T=0 proton--neutron pairing strength  $G^{T=0}_{pn}$
as function
of the deformation parameter $\beta_2$ for $^{64}Ge$, $^{68}Ge$ and
$^{76}Ge$.
}
\label{fig.5}
\end{figure}

It is an open issue  whether the value of
pairing strength $G^{T=0}_{pn}$ depends on the deformation
of the considered isotope.
In Fig. \ref{fig.5} this point is analyzed for
$^{64}Ge$, $^{68}Ge$ and $^{76}Ge$ assuming different deformations. The
$G^{T=0}_{pn}$ is displayed as a function of the
deformation parameter $\beta_2$
within the range $-0.4 \le \beta_2 \le 0.4$.
We see that $G^{T=0}_{pn}$ is sensitive to the change
of the quadrupole parameter $\beta_2$ especially if
the shape of the considered nucleus is oblate.
From the considered Ge isotopes $^{68}Ge$ exhibits
the strongest sensitivity of $G^{T=0}_{pn}$ to
$\beta_2$ parameter.

\begin{figure}
\begin{center}
\begin{tabular}{cc}
{\epsfig{figure=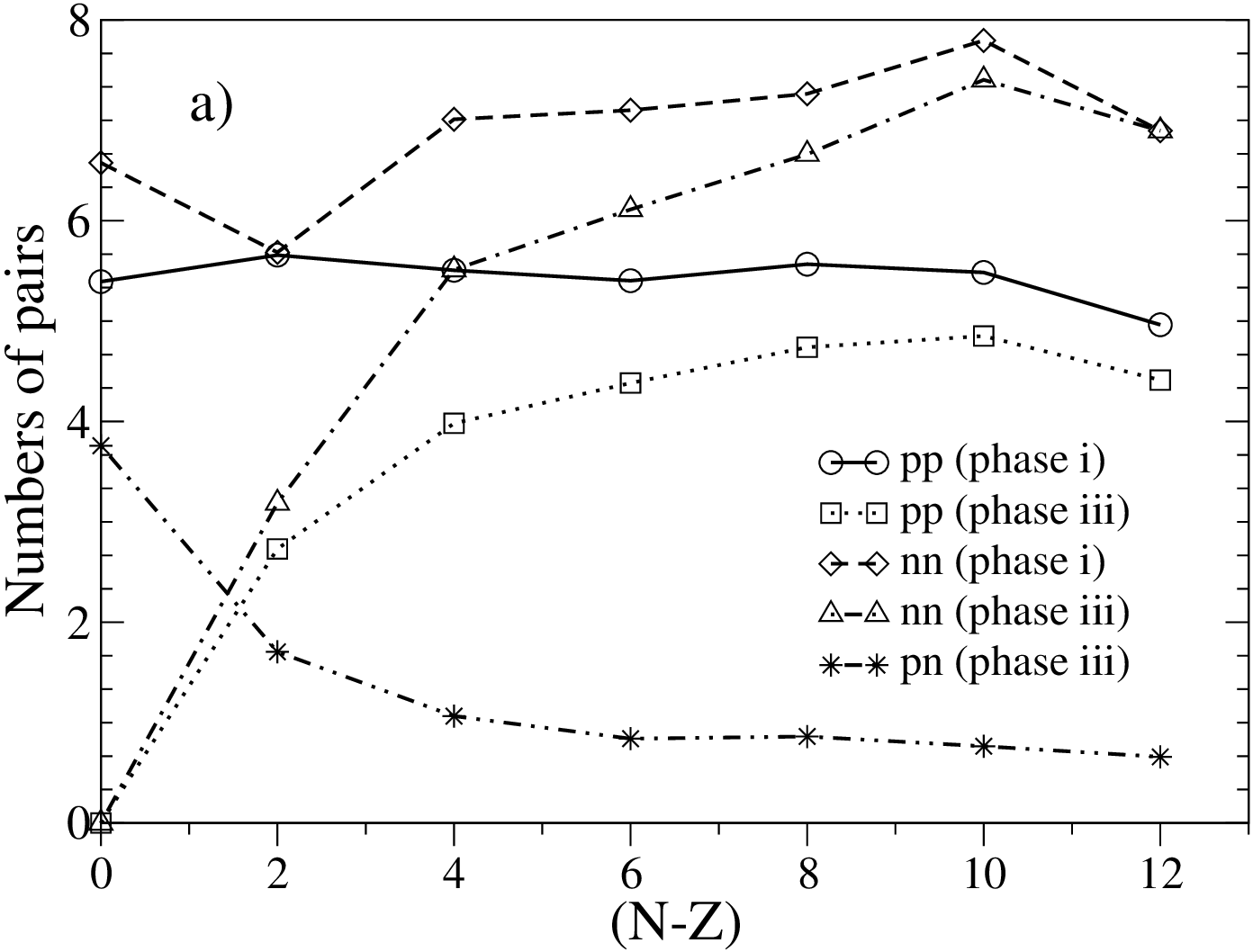,width=10cm} }
\\
\vspace{0.2cm}
\\
{\epsfig{figure=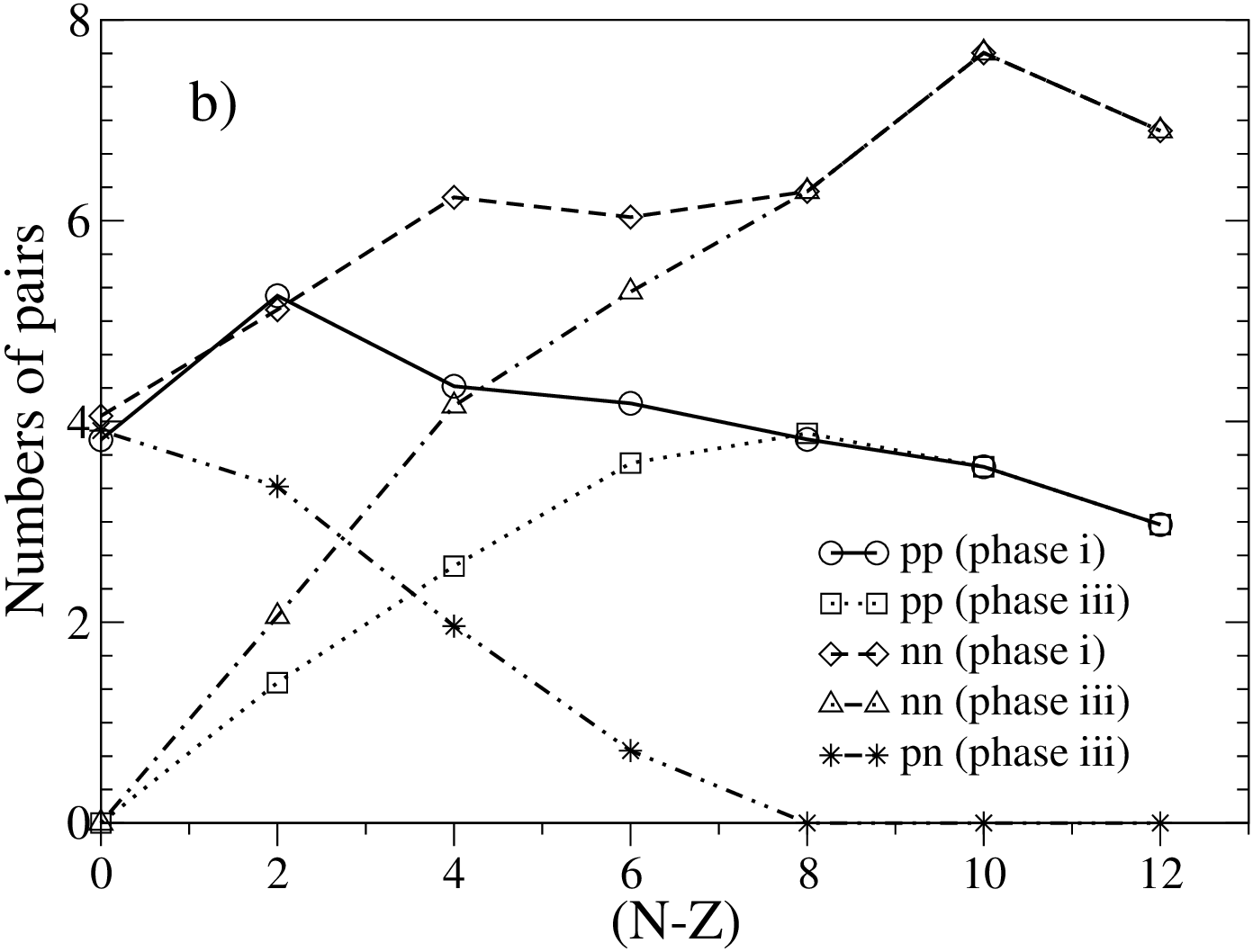,width=10cm} }
\end{tabular}
\end{center}
\caption{ The quantities $<{\cal N}_{pp}>$, $<{\cal N}_{nn}>$
and $<{\cal N}^{T=0}_{pn}>$ [representing number of pp,nn and pn
pairs; see Eqs. (\protect\ref{eq:13}) for
definition] for Ge isotopes, as a function
of $N-Z$. The results are presented for a pure like particle pairing
phase (phase i) and for  a phase where like-particle and T=0
proton-neutron pairs coexist (phase iii). The upper figure a)
refers to calculation with pairing strengths adjusted to experimental
pairing gaps and proton-neutron interaction energy. The lower figure b)
refers to calculation with pairing strengths given in Eq.
(\protect\ref{eq:18}).
}
\label{fig.6}
\end{figure}

In Fig. \ref{fig.6} the competition among pp, nn and pn pairs
in the ground state of even-even Ge isotopes is studied, as a
function of $N-Z$. The displayed quantities $<{\cal N}_{pp}>$,
$<{\cal N}_{nn}>$ and $<{\cal N}^{T=0}_{pn}>$
correspond roughly to the number of pp, nn and T=0 pn pairs
[see Eq. (\ref{eq:13})], respectively.
These quantities, as it was already stressed above, are
closely related to the different  contributions
to the total pairing energy (\ref{eq:11}). The number of pairs were measured
both for the  system with only like--particle pairs [phase i]
and for the system where like--particle and proton--neutron
pairs coexist [phase iii]. In Fig \ref{fig.6} a) the results
obtained with pairing strengths adjusted to experimental pairing gaps
($\Delta^{emp.}_{p}$ and $\Delta^{emp.}_{n}$)
and proton--neutron interaction energy ($\delta^{emp.}_{pn}$) are presented.
We see that in phase i) there is a rough constancy of the number of
pp-pairs for  Ge isotopes and that the number of nn-pairs is a little
bit larger and exhibits some fluctuations. There is a different
situation if the system of nucleons prefers the phase iii).
The $<{\cal N}_{pp}>$ and $<{\cal N}_{nn}>$ are equal to zero
for $^{64}Ge$ and grow up to maximum values about 7.6 and 4.8,
respectively, for $^{74}Ge$. We note that the behavior of
$<{\cal N}^{T=0}_{pn}>$ is different. The effect of the
proton-neutron pairing decreases  with increasing N-Z.
For large N-Z the value of $<{\cal N}^{T=0}_{pn}>$ is
significantly smaller as  $<{\cal N}_{pp}>$ and $<{\cal N}_{nn}>$,
but not negligible. If pairing strengths
given in (\ref{eq:18}) are used in the BCS calculation, one finds that the effect
of proton-neutron pairing disappears at $N-Z\geq 8$ in real
nuclei as it is shown in Fig. \ref{fig.6} b). Then, for these isotopes
one fails to explain non-zero value of the
proton-neutron interaction energy $\delta^{emp.}_{pn}$
(see Table \ref{table.1}). The values
of  $\delta^{emp.}_{pn}$ for all $^{70,72,74,76}Ge$
isotopes are of the same order.
Thus it is expected that the role of the pn-pairing
for all these isotopes is of comparable importance
and not negligible.

From the above discussion it follows that
the T=0 proton--neutron pairing correlations should be
considered also for medium-heavy nuclei with large neutron
excess, i.e., within a procedure proposed in this paper.
Usually, correlations between protons and neutrons in 
medium and heavy nuclei were neglected on the ground that 
two Fermi levels are apart. Here, it is shown that the 
proton-neutron pairing effect is not negligible for such
nuclear systems. 
We strongly suspect that the competition between the 
different kinds of pairs can affect measurable properties
of nuclei, in particular $\beta^+$ strengths.
The previous $\beta$- and $\beta\beta$-decay  studies
\cite{Cheoun95} performed within the spherical QRPA with T=1 
proton-neutron pairing support this conclusion as well.

\section{Summary and Conclusions}

We performed generalized BCS calculation by assuming axial symmetry
and the Hamiltonian with schematic T=1 and T=0
pairing forces in Eq. (\ref{eq:1}).
The system of BCS equations allows three different solutions.
There is one solution with only like--particle pairs, and two solutions
in which like and unlike particle pairs coexist, first with
T=1 and second with T=0 pn-pairs. We note that none of the observed
pairing modes allows simultaneous presence of both T=0 and T=1 
pn--correlations. The type of the pn pairs is determined by the
stronger from T=0 and T=1 pn pairing interactions of nuclear
Hamiltonian. For N=Z
$^{64}Ge$ pure T=0 pairing mode is found and a sharp
phase transition from the like particle pairing mode to 
unlike particle pairing mode is observed, 
what seems to be a result of a simple monopole pair Hamiltonian.
For pair Hamiltonians which are more complex, there is phase coexistence
between T=1 and T=0 pairing in N=Z nuclei, and not a sharp
transition from one to the other \cite{Goodman98,Goodman01}.
For other Ge isotopes the phase transition between different pairing
modes is much smoother.

A competition between like-particles and proton--neutron pairing
were studied in even-even Ge isotopes. The pairing strengths were
adjusted to reproduce the experimental odd-even mass differences.
The diminishing role of the pn pairs with increasing N-Z were
shown, however, the effect of proton neutron pairing were found
to be important also for isotopes with large neutron
excess N-Z, in particular for $^{76}Ge$, which undergoes double
beta decay. These results contrast with  the general belief that
proton-neutron pairing correlations are
restricted only to the vicinity of N=Z line. The values of the
calculated proton--neutron interaction energy $\delta^{emp}_{pn}$
for $N>Z$ isotopes are suggestive and should motivate a greater effort to
understand different properties of nuclei in the presence of the
T=0 proton-neutron pairing correlations.  However, we point out that
there is some disagreement between the calculation with
pairing strengths adjusted  to the experimental pairing gaps and
proton-neutron interaction energy and with commonly used
prescription for pairing strengths given in Eq. (\ref{eq:18}). 
Within the second scenario the deformed BCS solution with T=0 pairing 
was not found for $N-Z\ge 8$.

Of course the effect of pn-pairing on ground state properties of deformed 
nuclei can be studied self-consistently by solving the HFB equations
\cite{fae87}. In this paper we used advantage of the deformed BCS approach 
to estimate the effect of pn-pairing for $N>Z$ nuclei, which
can undergo single or double beta decay. Presently, there is a great
effort to increase the accuracy and reliability of the calculated single and 
double beta decay matrix elements. The effects of pn-pairing and deformation
on these matrix elements can be studied within a coupled deformed BCS plus 
QRPA approach \cite{beta03,subm}. The results of our paper indicate that some
of the beta and maybe also the  double beta decay observables might be influenced 
by the T=0 proton-neutron pairing.   

This work was supported in part by the Deutsche
Forschungsgemeinschaft (436 SLK 17/298), by the
``Land Baden-W\"urtemberg'' as a ``Landesforschungsschwerpunkt''
(III 1.3-H3-23/74/79/2003 ):
Low Energy Neutrinos'' and by the 
VEGA Grant agency of the Slovac Republic under contract
No. 1/0249/03.



\begin{table}
\caption{ The empirical [see Eq. (\protect\ref{eq:15})] and average
[see Eq. (\protect\ref{eq:16})]
pairing gaps and proton--neutron residual energy for Ge isotopes
with A=64-76.
}
\begin{tabular}{lccccccc}
 & \multicolumn{3}{c}{Empirical values} &
\multicolumn{4}{c}{Average values} \\ \cline{2-4}\cline{5-8}
 & & & & \multicolumn{2}{c}{Traditional m.} &
\multicolumn{2}{c}{Vogel et al.} \\ \cline{5-6} \cline{7-8}
Nucleus & $\Delta_{p}^{emp.}$ & $\Delta_{n}^{emp.}$ &  $\delta_{pn}^{emp.}$ &
 ${\overline{\Delta}}_{p,n}$ & ${\overline{\delta}}_{pn}$ &
 ${\overline{\Delta}}_{p,n}$ & ${\overline{\delta}}_{pn}$ \\ \cline{2-8}
 & [MeV] & [MeV] & [MeV] & [MeV] & [MeV] & [MeV] & [MeV]  \\\hline
$^{64}Ge$ & 1.807 & 2.141 & 1.498 & 1.500  & 0.313 & 1.800   & 0.484 \\
$^{66}Ge$ & 1.586 & 1.859 & 0.816 & 1.477 & 0.303 & 1.770  & 0.470 \\
$^{68}Ge$ & 1.609 & 1.882 & 0.630 & 1.455 & 0.294 & 1.727 & 0.455 \\
$^{70}Ge$ & 1.551 & 1.866 & 0.594 & 1.434 & 0.285 & 1.668 & 0.443 \\
$^{72}Ge$ & 1.614 & 1.836 & 0.583 & 1.414 & 0.278 & 1.600 & 0.430 \\
${^{74}Ge}$ & 1.621 & 1.715 & 0.424 & 1.350 & 0.270 & 1.523 & 0.419 \\
${^{76}Ge}$ & 1.561 & 1.535 & 0.388 & 1.376 & 0.263 & 1.441 & 0.408 \\
\end{tabular}
\label{table.1}
\end{table}



\newpage


\newpage


\newpage


\newpage


\newpage


\end{document}